\documentclass[CEJP,DVI]{cej} 
\usepackage{layout}
\usepackage{amsmath}
\usepackage{textcomp}
\usepackage{natbib}
\title
{
An adjustable law  of  motion  for  relativistic spherical
shells
}

\articletype{Research Article}

\author{ L. Zaninetti \inst{1}\email{E-mail: zaninetti@ph.unito.it} ,
            }
\institute{
 Dipartimento  di Fisica Generale,\\
Universit\`a degli Studi di Torino    \\
via P.Giuria 1,  10125 Torino,Italy    \\
}
\abstract
{
A  classical and a relativistic  law of motion
for an advancing shell are deduced
applying the thin layer approximation.
A  new  parameter connected with the quantity
of absorbed matter in the expansion   is introduced;
this  allows of  matching theory  and observation.
}
\def\snr{SN\,1993J\,}
\def\aap{A\&A\,  }
\def\apj{ApJ\,  }
\def\apjs{ApJS  }
\def\apss{Astrophysics and Space Science  }

\def\mnras{MNRAS\,  }

\begin{document}
\maketitle

\section{Introduction}

The first law  of motion  to be analyzed is
the  Sedov--Taylor  solution
\begin{equation}
R(t)=
\left ({\frac {25}{4}}\,{\frac {{\it E}\,{t}^{2}}{\pi \,\rho}} \right )^{1/5}
\quad,
\label{sedov}
\end{equation}
where $E$ is the energy injected into the process,
$\rho$  is  the density of the surrounding medium,
and $t$ is  the time,
see~\cite{Taylor1950a,Taylor1950b,Sedov1959}.
This equation  allows  of deducing the energy
of the first explosion of the  atomic bomb,
the Trinity test in New Mexico in 1945,
which  is  $E\approx 10^{21}$ ergs.
A second  application  of this equation
belongs to astrophysics and  is  the supernova explosion
(SN)  in which  $E\approx 10^{51}$ ergs,
see \cite{Dalgarno1987}.
A different approach as given
by \cite{Chevalier1982a} and \cite{Chevalier1982b}
analyzes  two  self-similar solutions
with varying inverse power law exponents
for the density profile of the advancing matter,
$R^{-n}$, and ambient  medium,
$R^{-s}$. The  previous assumptions give
a law  of motion
$R \propto  t^{\frac{n-3}{n-s} }$  when
 $n \, > 5$.
Another example  is an analytical solution suggested
by  \cite{Truelove1999} where the radius--time relationship
is regulated by the decrease in density:
as an example, a density  proportional to $R^{-9}$
gives  $R\propto t ^{2/3}$:
this mechanism as been applied to young
supernova remnants, see \cite{Vigh2011}.
The case of spherically symmetric thin shells  has  been
recently analyzed  both in special  relativity,
see  \cite{Gaspar2011} with application
to the mass inflation phenomenon, and
in general  relativity, see \cite{Kijowski2009}
where the solutions are split into  ``ordinary''
matter and  ``exotic'' matter.
The actual theoretical  situation
leaves a series of questions
unanswered or only partially answered:
\begin{itemize}
\item Is it possible to deduce an equation of motion for an
expanding shell by  assuming that only a fraction of the mass
enclosed in the advancing sphere is absorbed in the thin layer?
\item Can  this new adjustable  law of motion for an expanding
shell be calibrated on the observed  data? \item Can we deduce an
asymptotic behavior for the advancing radius of the thin layer of
the type $\propto t^{\alpha}$, where $t$ is the time and  $\alpha$
an observational exponent? \item Can we model the momentum carried
away by the photons? \item Can we derive a relativistic law of
motion for the expanding shell  under the previously outlined
hypothesis?
\end{itemize}
In order to answer these questions,
Section \ref{sec_classical}
reviews the existing situation and reports a
new classical
law of motion which  introduces
the two concepts
of porosity  and photon's losses.
Section \ref{sec_relativistic}
reviews a first relativistic law  of motion
and introduces a new
relativistic law of motion  which  includes
the  porosity.

\section{Classical  thin  layer  approximation}

\label{sec_classical}
This Section  reviews  the  general purpose
power  law  model, the standard thin layer
approximation with constant  density and the
thin layer approximation in a medium with density
regulated by a power law.
A first new equation of motion is introduced which
works  in a medium  with  constant density
but swept mass regulated  by a parameter called
porosity.
A second new equation of motion includes  porosity
and photon's losses.

\subsection{The  power law solution}
\label{powerlawsolution}
The equation of   the expansion  of a shell
can  be modeled by a power law  of the type
\begin{equation}
R(t) = R_0 (\frac{t}{t_0})^{\alpha}
\label{rpower}
\quad ,
\end{equation}
where
$R$ is the radius  of the expansion,
$t$ is the time,
$R_0$ is the radius  at  $t=t_0$,
and  $\alpha$ is an exponent which
can be found from a
numerical analysis.

The velocity is
\begin{equation}
v(t) = \alpha R_0 (\frac{1}{t_0})^{\alpha} t^{(\alpha-1)}
\quad .
\label {vpower}
\end{equation}

\begin{figure}
\begin{center}
\includegraphics[width=10cm]{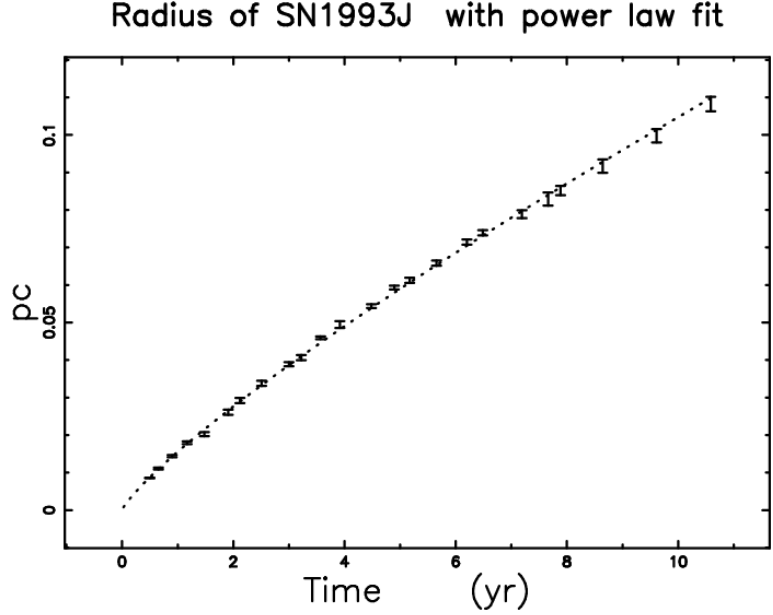}
\end {center}
\caption
{
Theoretical radius as given by the power law fit
represented by equation~(\ref{rpower})
with $\alpha$ =0.828, $R_0$=0.0087 pc and
$t_0$ = 0.498 yr.
The  astronomical data of \snr  are represented by
vertical error bars and
are extracted  from Table 1 in \cite{Marcaide2009}.
}
\label{1993pc_fit_power}
    \end{figure}

As an example, Fig. \ref{1993pc_fit_power}
reports the fit of
\snr where
the radius  is growing  more slowly
than the free expansion,
$R\propto\,t$,  and  more quickly than the
Sedov--Taylor  solution, $R \propto  t^{0.4}$,
see  equation (\ref{sedov}).

\subsection{Standard case}

\label{thinlayer}
The thin layer approximation assumes that all the swept-up
gas accumulates infinitely in a thin shell just after
the shock front.
The conservation of the radial momentum requires that
\begin{equation}
\frac{4}{3} \pi R^3 \rho \dot {R} =
\frac{4}{3} \pi {R_0}^3 \rho \dot {R_0}
\quad ,
\end{equation}
where $R$ and $\dot{R}$   are  the radius and the velocity
of the advancing shock,
$\rho$ the density of the ambient medium,
$R_0$ the initial radius
evaluated at $t=t_0$,
and
$\dot {R_0}$  the  initial velocity,
see \cite{Dyson1997,Padmanabhan_II_2001}.
The law of motion is
\begin{equation}
R = R_0 \left  ( 1 +4 \frac{\dot {R_0}} {R_0}(t-t_0) \right )^{\frac{1}{4}}
\label{radiusmthin}
\quad .
\end{equation}
and the velocity
\begin{equation}
\dot {R} = \dot {R_0} \left ( 1 +4 \frac{\dot {R_0}} {R_0}(t-t_0)\right )^{-\frac{3}{4}}
\label{velocitymthin}
\quad .
\end{equation}
From equation (\ref{radiusmthin}) we can extract $\dot {R_0}$
and insert it in equation (\ref{velocitymthin})
\begin{equation}
\dot {R} =\frac{1}{4(t-t_0)}  \frac{R^4-R_0^4}{R_0^3}
\left  ( 1+\frac{R^4-R_0^4}{R_0^4} \right )^{-\frac{3}{4}}
\label{velocitym2thin}
\quad .
\end{equation}
The astrophysical radius in pc as a function
of time is
\begin{equation}
R(t) =
\sqrt [4]{{{\it R_{0}}}^{3} \left(
 4.08\,10^{-6}\,{\it v_{1}}\, \left( t_1-t_{{0}} \right) +{\it R_{0}
} \right) } \, pc
\quad ,
\end{equation}
where
$t_1$ and $t_0$ are  times in years,
$R_0$ is the  radius in pc  when  $t_1=t_0$ and
$v_{1}$ is the velocity in
$\frac{1\,km}{s}$ units  when
$t_1=t_0$.
The momentum solution in the presence of a
constant density medium
scales as $t^{0.25}$.
More details can be found in \cite{Zaninetti2009a}.

\subsection{Classical medium with power law   }

\begin{figure*}
\begin{center}
\includegraphics[width=10cm]{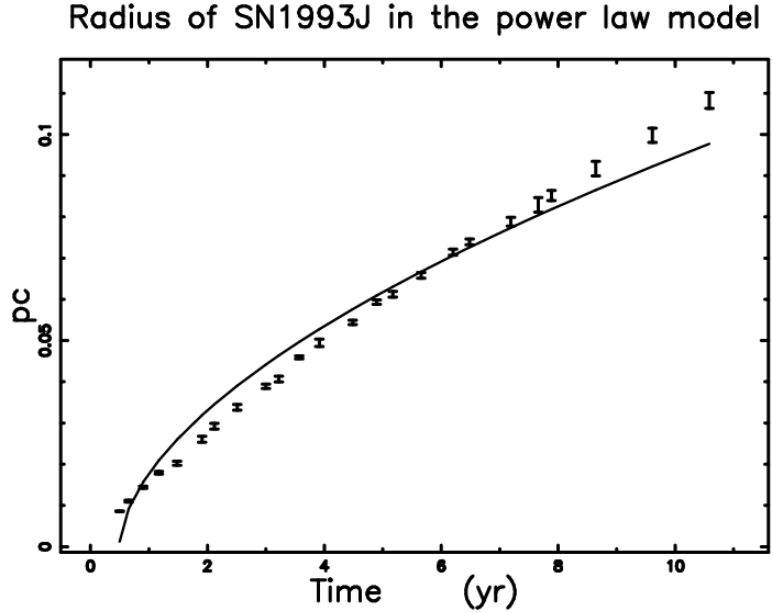}
\end {center}
\caption
{
Theoretical radius in pc as  function of the
time in yr  as obtained
by the solution of the non linear
equation (\ref{nonlineard})
(full line) which models the medium with variable density.
The  parameters of the simulation are
$d=2.24$  ,
$R_{0}$ = $0.031$ pc,
$t_{0}$ = $0.049$ ~yr
and $v_{0,1}$ =50 000.
The astronomical data of
\snr at  6 cm are
represented through
the  error bars    and are extracted from
Table 1 in \cite{Marcaide2009}
}
\label{fit_powermedium}
    \end{figure*}

We assume that around the
SNR the density of the
interstellar medium (ISM)
has
the following two piecewise dependencies
\begin{equation}
 \rho (R)  = \left\{ \begin{array}{ll}
            \rho_0                      & \mbox {if $R \leq R_0 $ } \\
            \rho_0 (\frac{R_0}{R})^d    & \mbox {if $R >    R_0 $ ~.}
            \end{array}
            \right.
\label{piecewiser}
\end{equation}
In this framework, the density decreases as
an inverse power law with an exponent $d$
which can be fixed from the
observed temporal evolution of the radius.
The mass swept, $M_0$,
in the interval $0 \leq r  \leq R_0$
is
\begin{equation}
M_0 =
\frac{4}{3}\,\rho_{{0}}\pi \,{R_{{0}}}^{3}
\quad .
\end{equation}
The mass swept ,$ M $,
in the interval $0 \leq r \leq R$
with $r \ge R_0$
is
\begin{eqnarray}
M =
-4\,{r}^{3}\rho_{{0}}\pi \, \left( {\frac {R_{{0}}}{r}} \right) ^{d}
 \left(d -3 \right) ^{-1}   \nonumber \\
+4\,{\frac {\rho_{{0}}\pi \,{R_{{0}}}^{3}}{d-
3}}
+ \frac{4}{3}\,\rho_{{0}}\pi \,{R_{{0}}}^{3}
\quad .
\end{eqnarray}
Momentum conservation in the thin
layer approximation
requires  that
\begin{equation}
M v = M_0 v_0
\quad ,
\end {equation}
where  $v$   is  the velocity at $t$
and    $v_0$ is  the velocity at $t=t_0$.
The previous expression as a function
of the radius
is
\begin{equation}
v  =
\frac
{
{{\it r_0}}^{3}{\it v_0}\, \left( 3-d \right)
}
{
3\,{{\it r_0}}^{d}{R}^{3-d}-{{\it r_0}}^{3}d
}
\quad .
\label{velclassicpowermedium}
\end{equation}
In this differential equation of first order
in $R$, the
variables can be separated and an integration
term-by-term gives
the following
nonlinear equation ${\mathcal{F}}_{NL}$
\begin{eqnarray}
{\mathcal{F}}_{NL} =
 \left( 4{R_{{0}}}^{3}d-{R_{{0}}}^{3}{d}^{2} \right) R -3{R_{{0}}}^
{d}{R}^{4-d}+{R_{{0}}}^{4}{d}^{2}
\nonumber \\
+12{R_{{0}}}^{3}v_{{0}}t+3{R_{{0}
}}^{4}-4{R_{{0}}}^{4}d
\nonumber\\
+7{R_{{0}}}^{3}v_{{0}}d{\it t_0}+{R_{{0}}}^{3
}v_{{0}}{d}^{2}t
\nonumber \\
-7{R_{{0}}}^{3}v_{{0}}dt
-12{R_{{0}}}^{3}v_{{0}}{
\it t_0}-{R_{{0}}}^{3}v_{{0}}{d}^{2}{\it t_0}
=0
\quad  .
\label{nonlineard}
\end {eqnarray}
An approximate solution of
${\mathcal{F}}_{NL}(r) $
can be obtained
assuming that  \\
$3 R_0^d R^{4-d}$
$\gg$
$-(4 R_0^3 d-R_0^3 d^2)R$
\begin{eqnarray}
 R(t) = \nonumber \\
 ( {R_{{0}}}^{4-d}-\frac{1}{3}d{R_{{0}}}^{4-d}
( 4-d ) \nonumber \\
 + \frac{1}{3}
 ( 4-d ) v_{{0}}{R_{{0}}}^{3-d} ( 3-d )
 ( t-t_{{0}} )  ) ^{\frac{1}{4-d}}
\quad .
\label{asymptotic}
\end{eqnarray}
Up to now, the physical units have not been specified: $pc$ for
length  and  $yr$ for time are the hybrid units  usually adopted
for SNRs. With these units, the initial velocity $v_{{0}}$ is
expressed in $\frac{pc}{yr}$ and should be converted into
$\frac{km}{s}$. This means that   $v_{{0}} =1.02\,10^{-6}
v_{{0,1}}$ where  $v_{{0,1}}$ is the initial velocity expressed in
$\frac{km}{s}$; in this transformation we have used
$1yr=3.15\times 10^{7}\;sec$ and $1~pc= 3.086\times10^{18} cm $.
The astrophysical version of the above equation
in pc is
\begin{eqnarray}
 R(t) = \nonumber \\
 ( {R_{0}}^{4-d}-\frac{1}{3}d{R_{{0}}}^{4-d}
( 4-d )
 +3.402 \,10^{-7} \times \nonumber \\
 \times( 4-d ) v_{{1}}{R_{{0}}}^{3-d} ( 3-d )
 ( t_1-t_{{0}} )  ) ^{\frac{1}{4-d}} \, pc
\quad ,
\label{radiusvarpc}
\end{eqnarray}
where
$t_1$ and $t_0$ are  times in  years,
$R_0$ is the radius in pc  at $t_1=t_0$ and
$v_{1}$ is the velocity at
$t_1=t_0$
in $\frac{km}{s}$.
The approximate solution (\ref{asymptotic})
has the
following limit  as $t \to \infty$
\begin{equation}
 R(t) =C_{th} t^{\frac{1}{4-d}}
\quad  ,
\end{equation}
where
\begin{equation}
C_{th} =
 \left( \frac{1}{3}\,{\frac {{{\it R_0}}^{3}{\it v_0}\, \left( -3+d \right)
 \left( -4+d \right) }{{{\it R_0}}^{d}}} \right)
^{ \frac {1} {\left( 4-d \right)}}
\quad .
\end{equation}
On imposing
\begin{equation}
\alpha =  { \frac {1} {\left( 4-d \right)}}
\quad  ,
\end{equation}
we obtain
\begin{equation}
d   =
{\frac {4\,\alpha-1}{\alpha}}
\quad  .
\end{equation}
where  $\alpha$ is an observable  parameter defined
in Section \ref{powerlawsolution}.
This means that the  unknown parameter
$d$ can be deduced
from  the observed parameter $\alpha$.
The numerical result is reported in
Fig. \ref{fit_powermedium}.
More details can be found in \cite{Zaninetti2011a}.

\subsection{Classical  porosity }

\label{sec_classicalporosity}
\begin{figure*}
\begin{center}
\includegraphics[width=10cm]{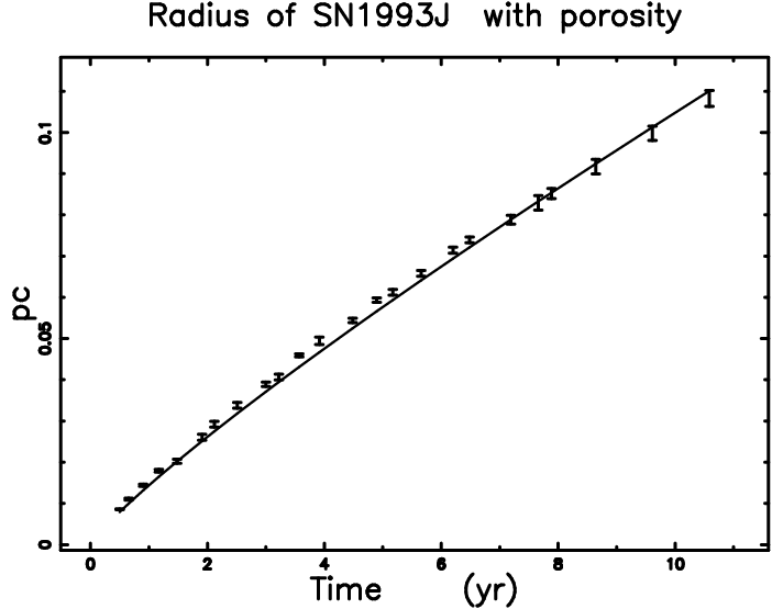}
\end {center}
\caption
{
Theoretical radius in pc as  function of the
time in yr  as obtained
by the solution of
Equation (\ref{rtclassicalporosity})
(full line) which models the medium with constant density
and porosity.
The  parameters of the simulation are
p=18.8 ,
$R_{0}$ = $5\,10^{-5}$ pc,
$t_{0}$ = $4.98\,10^{-6}$ ~yr
and $v_{0,1}$ =30 000.
The astronomical data of
\snr at  6 cm are
represented through
the  error bars    and are extracted from
Table 1 in \cite{Marcaide2009}
}
\label{fit_lev}
    \end{figure*}

The thin layer approximation
with porosity $p$
in classical physics
assumes that only a fraction
of the total mass enclosed
in the  volume
of the expansion
accumulates in a thin shell just after
the shock front.
The  global mass between  $0$ and $R_0$
is    $\frac {4}{3} \pi \rho  R_0^3  $
where
$\rho$ is the density of the ambient medium.
The mass that resides beyond the expansion is
\begin{equation}
M_0 =( \frac {4}{3} \pi \rho  R_0^3)^{\frac{1}{p}}
\quad  .
\end{equation}
The  mass swept between  $0$ and $R$
is
\begin{equation}
M =( \frac {4}{3} \pi \rho  R^3)^{\frac{1}{p}}
\quad  .
\end{equation}
The  parameter $p$ defined in the interval $[0, 1]$ has
been introduced because  the observed radius time relationship in
\snr can be approximated by  a power law dependence of the type
$R\, \propto  t^{0.82}$ \cite{Marcaide2009}. The theory suggests
$R \propto R^{2/5} $ for the Sedov solution, see equation
(\ref{sedov}) or $R \propto R^{1/4} $  for the thin layer
approximation, see equation (\ref{radiusmthin}); therefore these
two models does not match with the observations. We  will see
later on how is  possible deduce $p$ from the observed data.
The conservation of radial momentum requires that,
after the initial  radius $R_0$,
\begin{equation}
M   v =
M_0 v_0
\quad,
\end{equation}
where $R$ and $v$   are  the
radius and velocity
of the advancing shock.
In classical physics,
the velocity as a function of radius
is
\begin{equation}
\label{velocityclassicalporosity} v=  v_0 (\frac
{R_0}{R})^{\frac{3}{p}} \quad,
\end{equation}
and  introducing
$\beta_0= \frac {v_0}{c} $
and
$\beta= \frac {v}{c} $
we obtain
\begin{equation}
\beta =  \beta_0 (\frac {R_0}{R})^{\frac{3}{p}}
\quad .
\label{eqnbetaporosity}
\end{equation}
The law of motion is
\begin{equation}
R(t) =
\left( {R_{{0}}}^{1+3\,{p}^{-1}}+ \left( 3+p \right) v_{{0}}{R_{{0}}}
^{3\,{p}^{-1}} \left( t-t_{{0}} \right) {p}^{-1} \right) ^{{\frac {p}{
3+p}}}
\quad,
\label{rtclassicalporosity}
\end{equation}
where $t$ is the time and $t_0$ is
the initial  time.
In classical physics,
the velocity as a function of time
is
\begin{eqnarray}
v  =  \frac{N}{D}                                 \\
where                             \nonumber  \\
N=\left( {{\it R_0}}^{1+3\,{p}^{-1}}+ \left( 3+p \right) {\it v_0}\,{{
\it R_0}}^{3\,{p}^{-1}} \left( t-{\it t_0} \right) {p}^{-1} \right) ^{{
\frac {p}{3+p}}}{\it v_0}\,{{\it R_0}}^{3\,{p}^{-1}}p
\nonumber \\
D={{\it R_0}}^{1+3\,{p}^{-1}}p+3\,{\it v_0}\,{{\it R_0}}^{3\,{p}^{-1}}t+{
\it v_0}\,{{\it R_0}}^{3\,{p}^{-1}}tp
\nonumber  \\
-3\,{\it v_0}\,{{\it R_0}}^{3\,{p}^{-
1}}{\it t_0}-{\it v_0}\,{{\it R_0}}^{3\,{p}^{-1}}{\it t_0}\,p
\, .
                        \nonumber \\
\label{velocitymporosity}
\end{eqnarray}
Equation (\ref{rtclassicalporosity}) can  also be solved
with a similar solution of type
$R=K(t-t_0)^{\alpha}$,
$k$ being a constant,
and the classical result is
\begin{equation}
\label{radiussimilarporosity}
R(t) =
 \left( \frac{  \left( 3+p \right) v_{{0}}{R_{{0}}}^{3/p}
\left( t-t_
{{0}} \right) }{p}  \right) ^{{\frac {p}{3+p}}}
\quad .
\end{equation}
The similar  solution  for the  velocity is
\begin {equation}
v(t) =
 \left( 3+p \right) ^{-\frac {3} { 3+p}}{p}
^{\frac {3}{3+p}}
{{\it v_0}}^{{\frac {p}{3+p}}}{{\it R_0}}
^{\frac{3}{3+p}}
\left( t-{\it t_0} \right) ^{-\frac {3} { 3+p}}
\quad  .
\label{velocitysimilarporosity}
\end{equation}
The similar formula (\ref{radiussimilarporosity})
for the radius  can be compared
with  the observed
radius--time  relationship
reported as
\begin{equation}
R(t) = r_{obs} t^{\alpha_{obs}}
\label{rpowerporosity}
\quad,
\end{equation}
where the two parameters $r_{obs}$
and $\alpha_{obs}$ are
found from the numerical analysis
of the observational
data, see  Section \ref{powerlawsolution}.
In this case the velocity  is
\begin{equation}
v(t) = r_{obs} \, \alpha_{obs} t^{(\alpha_{obs}-1)}
\quad .
\label {vpowerporosity}
\end{equation}
The comparison between
theory and observation
allows of deducing  $p$:
\begin{equation}
p  = \frac{3 \alpha_{obs}}{ 1 -\alpha_{obs}}
\quad   .
\end{equation}
The theoretical solution
as given  by Equation (\ref{rtclassicalporosity})
can be found through
the  Levenberg--Marquardt  method (subroutine
MRQMIN in \cite{press})
and  Fig.~\ref{fit_lev} reports
a numerical  example.

\subsection{Classical  porosity and photon losses }

\begin{figure*}
\begin{center}
\includegraphics[width=10cm]{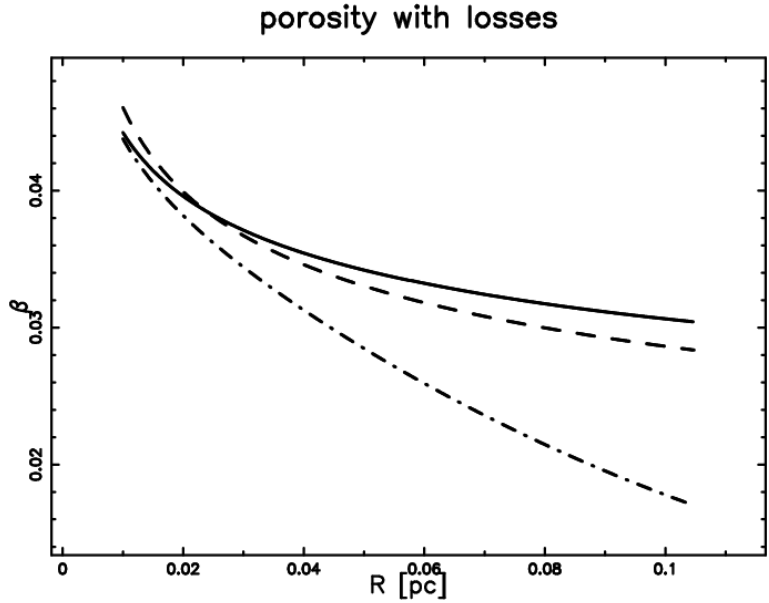}
\end {center}
\caption
{
Theoretical classical velocity with porosity
as  given  by Equation (\ref{velocityclassicalporosity})
(full  line),
velocity as  given  by the power law dependence
as in Equation (\ref{vpowerporosity}) (presence of porosity)
(dash-dash line) and
velocity with porosity and  photon losses as given by equation
(\ref{eqnvelocity_losses}) when $C_l$=50 (dash-dot-dash line).
The  parameters of the simulation are
p=18.8 ,
$R_{0}$ = $5\,10^{-5}$ pc,
$t_{0}$ = $4.98\,10^{-6}$ ~yr
and $v_{0,1}$ =30 000.
}
\label{velocity_losses}
    \end{figure*}

\begin{figure*}
\begin{center}
\includegraphics[width=10cm]{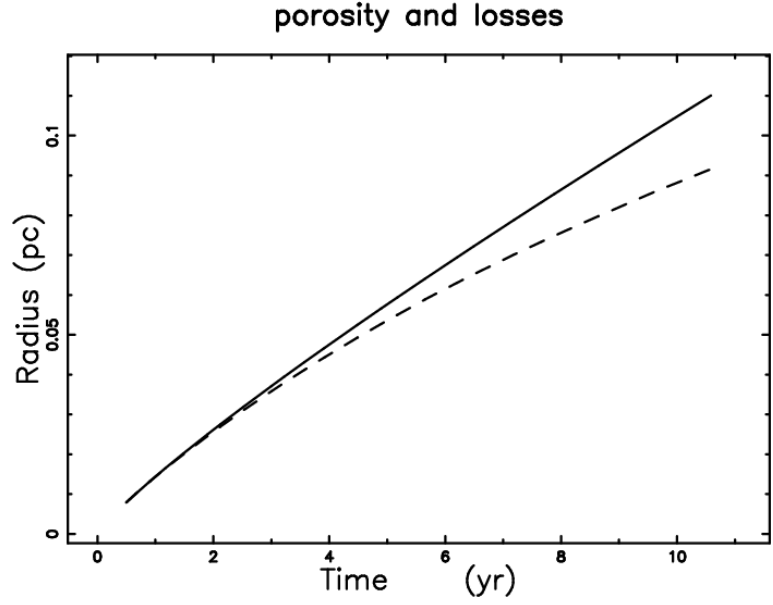}
\end {center}
\caption
{
Theoretical radius in pc as  function of the
time in yr  as obtained
by the solution of
Equation (\ref{rtclassicalporosity})
with porosity
(full     line)
and  radius  with with porosity  and  losses
as  given by Equation (\ref{eqnfit_lev_losses})
         (dashed   line)
when $C_l$=50.
The  parameters of the simulation are
$p$=18.8 ,
$R_{0}$ = $5\,10^{-5}$ pc,
$t_{0}$ = $4.98\,10^{-6}$ ~yr
and $v_{0,1}$ =30 000 .
}
\label{fit_lev_losses}
    \end{figure*}

The emission of photons
produces losses
in the momentum carried
by the advancing shell.
This effect  can be  parametrized by the following
conservation law
\begin{equation}
 \left( 4/3\,\rho\,\pi \,{R}^{3} \right) ^{{p}^{-1}}v+C_{{l}} \left( 4
/3\,\rho\,\pi  \right) ^{{p}^{-1}} \left( {R}^{2}-{R_{{0}}}^{2}
 \right) v
  = M_0 v_0
\quad ,
\end {equation}
where  $ C_{{l}} \left( 4
/3\,\rho\,\pi  \right) ^{{p}^{-1}} \left( {R}^{2}-{R_{{0}}}^{2}\right )  $
takes account of the losses
and $C_l$ is  a constant.
At  the  moment  of writing
is not possible  to  deduce the  parameter $C_l$
which controls  the quantity of momentum
carried away  by the photons due to the presence
of another parameter $p$ which controls
the  quantity of swept mass during the expansion.
The expression for the velocity  in
presence of porosity
and  momentum  losses by a
radiative process
is
\begin{equation}
\beta=
\frac
{
\left( {R_{{0}}}^{{p}^{-1}} \right) ^{3}\beta_{{0}}
}
{
\left( {R}^{{p}^{-1}} \right) ^{3}+C_{{l}}{R}^{2}-C_{{l}}{R_{{0}}}^{2
}
}
\label{eqnvelocity_losses}
\quad .
\end{equation}
Fig.  \ref{velocity_losses}
reports
how  a given value    of  $C_l$
modifies the velocity.
The  non linear equation  for the trajectory
in presence  of  porosity  and losses
is
\begin{eqnarray}
-{\frac { \left( {R_{{0}}}^{{p}^{-1}} \right) ^{3}R_{{0}}p}{3+p}}+2\,{
\frac {C_{{l}}{R_{{0}}}^{3}}{3+p}}+2/3\,{\frac {C_{{l}}{R_{{0}}}^{3}p}
{3+p}}+{\frac { \left( {R}^{{p}^{-1}} \right) ^{3}Rp}{3+p}}+1/3\,{
\frac {C_{{l}}{R}^{3}p}{3+p}}
\nonumber  \\
+{\frac {C_{{l}}{R}^{3}}{3+p}}
-{\frac {C_
{{l}}{R_{{0}}}^{2}Rp}{3+p}}-3\,{\frac {C_{{l}}{R_{{0}}}^{2}R}{3+p}}-
 \left( {R_{{0}}}^{{p}^{-1}} \right) ^{3}\beta_{{0}}c \left( t-{\it t0
} \right) =0
\label{eqnfit_lev_losses}
\quad  .
\end{eqnarray}
An example  of the  modification of the  trajectory
as  given  by the parameter $C_l$  is  reported
in Fig.  \ref{fit_lev_losses}.

\section{Relativistic  thin  layer  approximation}

\label{sec_relativistic}

The Newton's second law in special relativity is:
\begin{equation}
F = \frac {dP} {dt } = \frac {d} {dt} ( m v)=
\frac {d} {dt} ( \frac{ m_0 v} {\sqrt{1-\frac{v^2}{c^2}}})
\quad ,
\label{newtonrelativistic}
\end{equation}
where $F$ is the force,
$P$ is the relativistic momentum,
$m$ is the relativistic mass,
$m_0$ is the mass at rest and
$v$ is the velocity,
see  equation~(7.16) in~\cite{French1968}.
This Section  reviews
the relativistic thin layer approximation in a medium
with density  regulated by a power law.
A  new relativistic equation of motion is introduced which
works  in a medium  with  constant density
but adjustable  swept mass.

\subsection{Relativistic power law}

\begin{figure*}
\begin{center}
\includegraphics[width=10cm ]{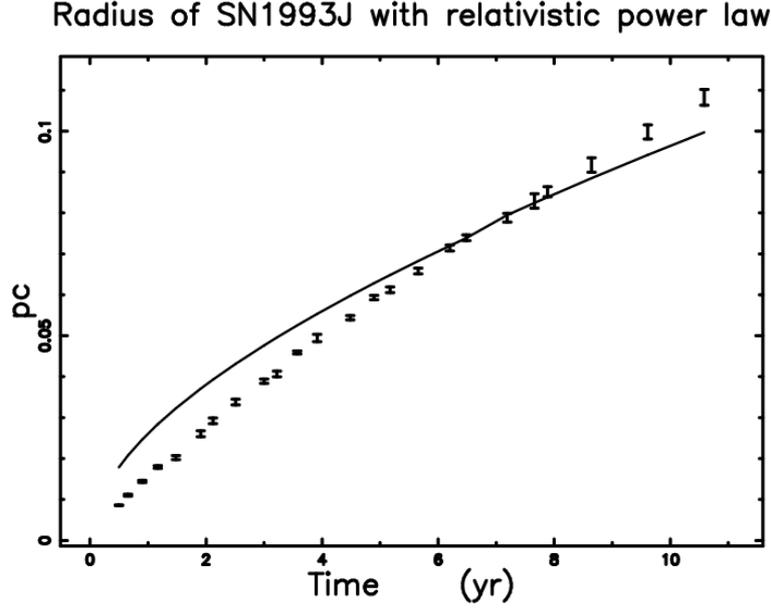}
\end {center}
\caption
{
Theoretical radius as obtained
by the solution of  the relativistic
equation (\ref{eqnrelpower})
which models the medium with variable density (full line).
The  parameters of the simulation are
$d=2.52$  ,
$R_{0}$ = $0.0068$ pc,
$t_{0}$ = $0.14$ ~yr     ;
and $\beta_0=0.30$.
The astronomical data of \snr are represented
with
vertical error bars.
}
\label{1993pc_fit_rel_power}
    \end{figure*}

In the case of the relativistic expansion of a shell
in which all the swept material
resides at two different points,
denoted by radius $ R $  and radius $R_0$,
the previous  equation (\ref{newtonrelativistic})
gives:
\begin{equation}
M \frac
{ \beta }
{\sqrt  {1-\beta^2}}
=
M_0 \frac
{\beta_0 }
{\sqrt  {1-\beta_0^2}}
\quad ,
\end{equation}
where
$\beta_0$=$v_0/c$,
$\beta$=$v/c$,
$M$   is the rest mass swept between  0 and $R$ and
$M_0$ is the rest mass swept between  0 and $R_0$.
This  formula is invariant
under Lorentz transformations and the initial
velocity, $v_0$, cannot be greater than the velocity of light.
Assuming a  spatial  dependence of the ISM
as given by formula~(\ref{piecewiser}),
the relativistic conservation of momentum
gives
\begin{equation}
\frac
{
-4\,\rho\,\pi \, \left( 3\,{r}^{3} \left( {\frac {{\it r_0}}{r}}
 \right) ^{d}-{{\it r_0}}^{3}d \right) \beta
}
{
3\, \left( -3+d \right) \sqrt {1-{\beta}^{2}}
}
=
\frac
{
4\,\rho\,\pi \,{{\it r_0}}^{3}{\it \beta_0}
}
{
3\,\sqrt {1-{{\it \beta_0}}^{2}}
}
\quad .
\end{equation}
According to the previous equation,
$\beta$ is
\begin{eqnarray}
\beta
=
\frac
{
- \left( -3+d \right) {\it \beta_0}\,{{\it r_0}}^{3}
}
{
\sqrt {D_{\beta}}
}
\label{velrelativisticpower}
\\
with~D_{\beta}=
9\,{R_{{0}}}^{6}{\beta_{{0}}}^{2}-6\,{R_{{0}}}^{6}{\beta_{{0}}}^{2}d
\nonumber \\
 +9 \,{r}^{6-2\,d}{R_{{0}}}^{2\,d}
\nonumber  \\
-6\,{r}^{3-d}{R_{{0}}}^{d+3}d+{R_{{0}}}^ {6}{d}^{2} \nonumber \\
-9\,{\beta_{{0}}}^{2}{r}^{6-2\,d}{R_{{0}}}^{2\,d}+6\,{\beta_
{{0}}}^{2}{r}^{3-d}{R_{{0}}}^{d+3}d \nonumber \quad  .
\end{eqnarray}
In this differential equation of
first order in $r$, the
variables can be separated and the integration
can be expressed as
\begin{equation}
\int_{R_0}^{R} \sqrt {D_{\beta}} dR
=
c \left( 3-d \right) \beta_{{0}}{R_{{0}}}^{3} \left( t-t_{{0}}
 \right)
\quad .
\label{eqnrelpower}
\end{equation}
The integral of the previous equation
can be performed analytically only in the
cases $d=0$, $d=1$ and $d=2$,
but we do not report the result because
we are interested in a variable value of $d$.
The integral can be easily evaluated
from a theoretical point of view
using the subroutine  QROMB
from \cite{press}.
The numerical result is reported in
Fig. \ref{1993pc_fit_rel_power}.
More details  can be found in
\cite{Zaninetti2010e,Zaninetti2011a}.

\subsection{Relativistic  porosity}

\label{sec_relativisticporosity}
\begin{figure*}
\begin{center}
\includegraphics[width=10cm]{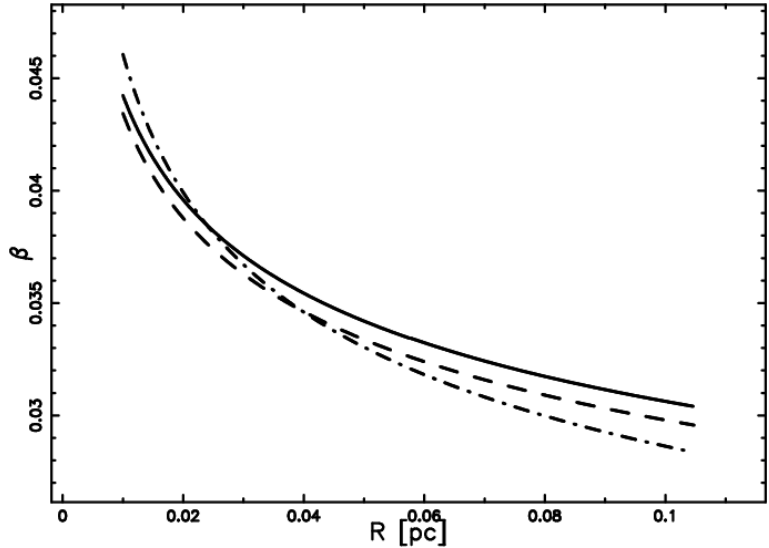}
\end {center}
\caption
{
Theoretical
relativistic velocity  for  \snr
as given by
Equation (\ref{velocityrelativisticporosity})
(relativistic porosity-dashed line),
theoretical classical velocity
as  given  by Equation (\ref{velocityclassicalporosity})
(classical porosity -full  line),
velocity as  given  by the power law dependence
as in Equation (\ref{vpowerporosity})
(dot-dash-dot-dash line).
}
\label{velocityporosity}
    \end{figure*}

\begin{figure*}
\begin{center}
\includegraphics[width=10cm]{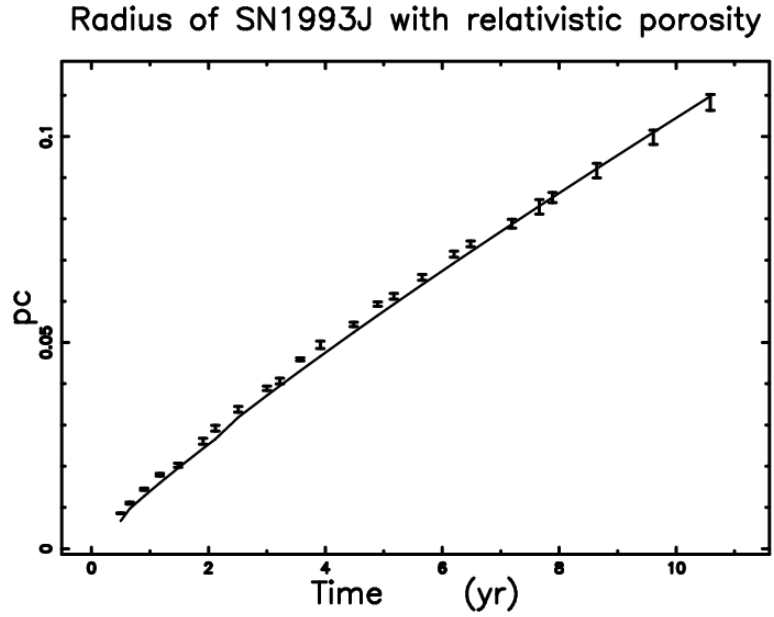}
\end {center}
\caption
{
Theoretical radius in pc as  function of the
time in yr
as obtained
by the solution of  the
relativistic
Equation (\ref{eqnrelporosity}) (relativistic  porosity)
(full line).
The  parameters of the simulation are
p=18.3,
$R_{0}$ = $6\,10^{-5}$ pc,
$t_{0}$ = $5\,10^{-8}$ ~yr,
and $\beta_0$ =0.1.
The astronomical data of
\snr are
represented through
error bars
and are extracted from
Table 1 in \cite{Marcaide2009}
}
\label{1993pc_fit_rel}
    \end{figure*}

\begin{figure*}
\begin{center}
\includegraphics[width=10cm]{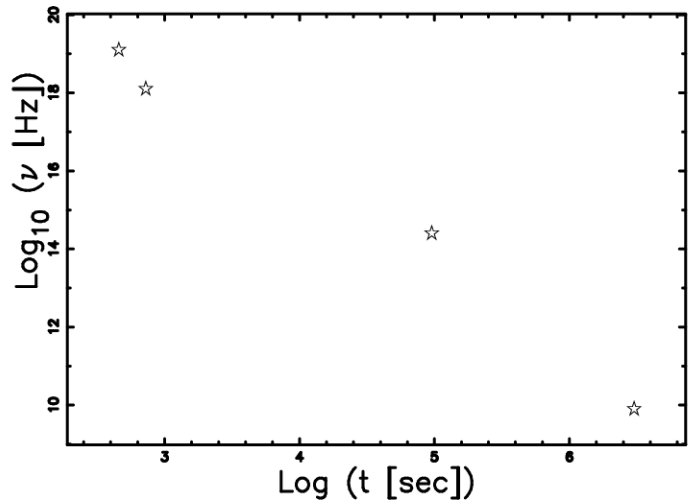}
\end {center}
\caption { Bands from the radio through the  IR/optical to X-ray
and BAT at  which  the expansion becomes optically thin. The
energy bands are expressed in Hz and the y-axis  as the logarithm
of the frequency. The  x-axis reports the logarithm of the time in
$s$. The data are extracted by the author from Table 1 in
\cite{Gou2007}. } \label{freq_time_grb}
    \end{figure*}

\begin{figure*}
\begin{center}
\includegraphics[width=10cm]{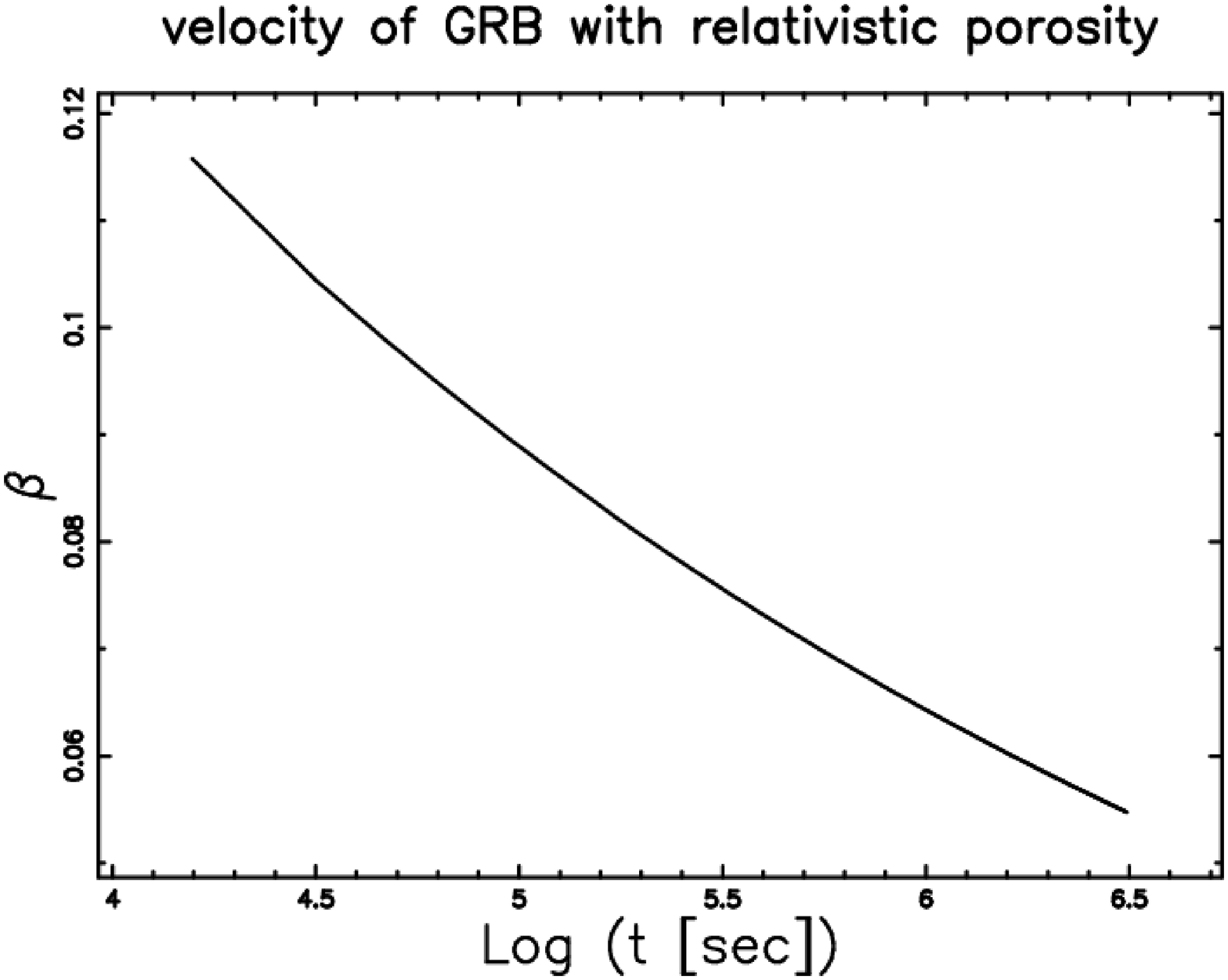}
\end {center}
\caption { Theoretical relativistic velocity  for  \snr as given
by Equation (\ref{velocityrelativisticporosity}) as  function of
Log(time) in seconds. The  parameters of the simulation are
p=18.3, $R_{0}$ = $2.25\,10^{-8}$ pc, $t_{0}$ = $5\,10^{-8}$ ~yr,
and $\beta_0$ =0.33. } \label{vel_grb}
    \end{figure*}

According to equation (\ref{newtonrelativistic})
the Newton's law in special relativity is
\begin{equation}
F = \frac {dp} {dt } = \frac {d} {dt} ( M v)=
\frac {d} {dt} ( \frac {M_r v}
{
\sqrt{1 -\frac{v^2}{c^2}}
}  )
=
\frac {d} {dt} ( \frac {M_r v}
{
\sqrt{1 -\beta^2}
}  )
\quad,
\end{equation}
where $F$ is the force,
      $p$ is the relativistic momentum,
      $M$ is the relativistic mass,
      $M_r$ is the mass at rest,
      $v$ is the velocity,
      $\beta=\frac{v}{c}$,
and
$c$ is  the speed of light.
In the case of the relativistic expansion
of a shell
in which  the swept material
is  part of the mass
contained in the advancing layer,
the equation of motion is
\begin{equation}
\frac
{(\rho \frac {4}{3} \pi R^3)^{1/p}    \beta }
{\sqrt  {1-\beta^2}}
=
\frac
{(\rho \frac {4}{3} \pi R_0^3)^{1/p}  \beta_0 }
{\sqrt  {1-\beta_0^2}}
\quad,
\end{equation}
where
$\beta_0$=$v_0/c$,
$R_0$ is the initial radius, and
$R$   is the temporary radius.
The velocity  of a relativistic expanding shell is
\begin{equation}
\label {velocityrelativisticporosity}
\beta=
\frac
{
\beta_{{0}}{R_{{0}}}^{3/p}
}
{
\sqrt {{R_{{0}}}^{6/p} {\beta_{{0}}}^{2}+{R}^{6/p}-{R}^
{6/p}{\beta_{{0}}}^{2}}
}
\quad .
\end{equation}
Fig.~\ref{velocityporosity} shows the classical
and relativistic
behaviors of the velocity as a
function of the radius $R$.
In this differential equation of
the first order in $r$ the
variables can be separated and the integration
can be expressed as
\begin{equation}
\int_{R_0}^{R}
\sqrt {{R_{{0}}}^{6/p}{\beta_{{0}}}^{2}+{R}^{6/p}-{R}^
{6/p}{\beta_{{0}}}^{2}}
 dr
=
\beta_{{0}}{R_{{0}}}^{3/p} (t-t_0) dt
\quad .
\label{eqnrelporosity}
\end{equation}
The integral of the
previous equation
can be expressed analytically
but the result is complicated.
The
integral can be easily evaluated numerically using the subroutine
QROMB from  \cite{press}.
The numerical result is reported in
Fig.~\ref{1993pc_fit_rel}.
An interesting property of the Gamma-ray burst  (GRB) is that the
frequency  of observation at which it starts to be visible
decreases with time. An example of such behavior for  GRB 050904
in the BAT, XRT, J, and I bands, can be found in Table 1 of
\cite{Gou2007}  . The numerical analysis of the data as reported
in Fig. \ref{freq_time_grb} gives
\begin{equation}
\nu = 1.03 10^{25} t(s)^{-2.27}\,  Hz \label{trasparente}
\end{equation}
where $t$  has  been chosen as the value of time at which the flux
at the chosen frequency  is  maximum.
The interval in time at which the GRB becomes visible in different
astronomical bands is here correlated with the velocity of the
shell's  expansion, see Fig. \ref{vel_grb}.

\section{Conclusions}

We have deduced two new laws of motion for an advancing shell
assuming that only a fraction of the mass  which  resides in the
surrounding medium  is  accumulated in the advancing layer.
According to the present  laws of mechanics the   analysis has
been split into classical and relativistic cases, see  Equations
(\ref{rtclassicalporosity}) and (\ref{eqnrelporosity}). In the
classical  case a similar  solution has  been  found, see Equation
(\ref{radiussimilarporosity}). The  momentum carried  away  by the
photons can be modeled in the classical case by introducing a
reasonable form for the  photon's losses , see  (
\ref{eqnfit_lev_losses}).
\section*{ Acknowledgements}
I would like to thank the two  anonymous referees for constructive
comments on the text.

\end{document}